%% file: main.tex
\documentclass{article}
\usepackage[utf8]{inputenc}

\title{NeurIPS Competition Instructions and Guide: Causal Insights for Learning Paths in Education}
\author{\textbf{Wenbo Gong}$^*$\thanks{*Microsoft Research}\and Digory Smith$^\dagger$\thanks{$^\dagger$Eedi}\and Zichao Wang$^\ddagger$\thanks{$\ddagger$Rice University}\and Craig Barton$^\dagger$ \and Simon Woodhead$^\dagger$\and Nick Pawlowski$^*$\and Joel Jennings$^*$\and Cheng Zhang$^*$}
\date{June 2022}
\renewcommand\footnotemark{}
\input{inputs}

\begin{document}
\maketitle

\begin{abstract}
In this competition, participants will address two fundamental {causal} challenges in {machine learning in the context of } education using time-series data. The first is to identify the causal relationships between different constructs, where a construct is defined as the smallest element of learning. The second challenge is to predict the impact of learning one construct on the ability to answer questions on other constructs. Addressing these challenges will enable optimisation of students' knowledge acquisition, which can be deployed in a real edtech solution impacting millions of students. Participants will run these tasks in an idealised environment with synthetic data and a real-world scenario with evaluation data collected from a series of A/B tests. 

\end{abstract}

\section{Introduction}
\input{Background}
\input{Impact}
\input{Novelty}
\input{Website}

\input{Data}
\input{DataRealWorld}

\input{Task}

\input{Submission_protocol}
\input{starting_kit}
\input{Organisation}

\input{Resources}

\bibliographystyle{apalike}
\bibliography{reference}
\end{document}

%% file: inputs.tex
\usepackage{amsmath, amsthm, amssymb}
\usepackage{bm}
\usepackage{natbib}
\usepackage{xcolor}
\usepackage[margin=1.3in]{geometry}
\usepackage{booktabs}       
\usepackage[utf8]{inputenc} 
\usepackage[T1]{fontenc}    
\usepackage{hyperref}       
\usepackage{url}            
\usepackage{amsfonts}       
\usepackage{nicefrac}       
\usepackage{microtype}      
\usepackage{ulem}
\usepackage{pdflscape}
\usepackage{afterpage}
\usepackage{caption}

\DeclareMathOperator{\cate}{CATE}
\DeclareMathOperator{\doop}{do}

%% file: Background.tex
Causal machine learning is a field that focuses on using machine learning method to tackle causality problems. Despite the recent progress of this field, there are still many unresolved challenges including
missing data, selection bias, unobserved confounders, etc., which are ubiquitous in the real world.  Advances in any of the above areas can greatly reduce the gap between research and real world impact.
In this competition, we focus on two fundamental challenges of causal machine learning in the context of education using time-series data.
The first is to identify the causal relationships between different constructs, where a construct is defined as the smallest element of learning.
The second challenge is to predict the impact of learning one construct on the ability to answer questions on other constructs.
Addressing these challenges will not only impact the causal ML community but also enable optimisation of students' knowledge acquisition, which can be deployed in a real edtech solution impacting millions of students.
Participants will run these tasks in an idealised environment with synthetic data and a real-world scenario with evaluation data collected from a series of A/B tests.
We expect participants to develop novel machine learning methodologies for causal discover between different constructs and the impact estimation of learning one construct on other constructs, 
which should bring fundamental advances to causal ML.
\paragraph{Background}
A significant challenge in education is deciding the order in which a subject should be taught, to allow personalised learning journeys. We can break down topics into constructs, which are the smallest element of learning. For example "Converting between cm and m" is a construct within the topic "Units of length". Discovering the relationships between constructs can help the design of more efficient curricula.

We cannot measure student's construct knowledge directly, so we use answers to questions as a surrogate measure. A deceptively simple but powerful question type used to understand student learning is a \textit{diagnostic question}. This is a multiple-choice question with four answers designed to assess a single construct. By inspecting the correctness to diagnostic questions from large number of students should reveal the hidden relationships behind different constructs (e.g. some constructs can be prerequisite for others), which can be leveraged to improve the students' knowledge acquisition.

The team behind Eedi (\url{https://eedi.com}) created a platform (\url{https://diagnosticquestions.com/}) to facilitate teachers crowd-sourcing these diagnostic questions. Students can complete quizzes and lessons on Eedi and this real-world data will be made available for this competition. 

%% file: Impact.tex
\paragraph{Impact on education}

In this competition, we challenge participants to develop novel methodologies to discover the causal relationships between different constructs and evaluate the impact of learning one construct on other constructs. 

The real-world impact would be to provide an objective construct map which acts as an architecture of learning. This could be used in many educational contexts, for example to order topics within curricula or to direct students to prerequisites. 

The real-world data for this competition comes from an educational platform that is already deployed and used at scale. The data describes real answers given by real students to real questions. By providing the opportunity to work on genuine educational data and real problems in an engaging manner, our competition will attract talent to the important field of machine learning in education.

We expect the competition to bring fundamental advances to educational data mining technologies. These methods will be deployed in a real educational platform where they will improve the learning outcomes of millions of students. 

\paragraph{Impact on the machine learning community}

Assumptions are important in the machine learning community. However, for real-world applications, e.g. the education domain, many of the common assumptions are not satisfied. This poses fundamental challenges: 
(i) In the education domain, incomplete observations are ubiquitous, which poses serious challenges to causal discovery and inference; (ii) The educational data may be discrete or collected using irregular time intervals, which violates common assumptions for many causal models;  (iii) The observational data can be collected with selection bias (e.g. students with higher grades tend to have higher course completion rates); (iv) It is impossible to measure all confounders in real life. How to handle latent confounders is still unsolved. 
   
Addressing these challenging problems within the educational domain will present interesting technical challenges and lead to significant contributions to NeurIPS and the broader machine learning community. First, we expect the users to develop novel models to tackle the aforementioned challenges presented in the data. These could include, but are not limited to, developing new constraint or score based time-series causal models that simultaneously incorporate missing mechanisms with irregular sampling time and selection bias. Second, in this challenge, we provide an opportunity for real-world evaluations of causal models with the interventional data from randomized A/B tests. Finally, causal techniques have been extensively used in economics, biology, health care, climate science, etc. However, they are less explored within the educational context, this challenge opens up a window for applying causal techniques to a new domain.

%% file: Novelty.tex
\paragraph{Novelty}
Our competition is novel in that there are very few competitions focusing on real-world evaluations of causal models. To the best of our knowledge, there are two previous causal competitions: (1) Causality for climate\footnote{https://causeme.uv.es/neurips2019/}; (2) Learning by doing\footnote{https://learningbydoingcompetition.github.io}. Our competition differs in terms of data and proposed tasks. Specifically, the data from both competitions are synthetic to some extent, whereas we include real-world data. In addition, competition (1) mainly focused on causal discovery tasks whereas ours also include inference tasks. 

Our proposed competition can be regarded as a sequel of our previously accepted NeurIPS competition \footnote{https://competitions.codalab.org/competitions/25449}. However, this competition also differs significantly from the previous one in terms of the proposed tasks, collected data and the final goal. In particular, data for the proposed competition is a time-series of student-answer pairs whereas the data in the previous competition was static. We provide both idealized synthetic data from a simulator and real-world data, which present different levels of difficulty. For evaluation, we collect the interventional data from both simulator and real-world A/B tests. The proposed tasks will explore the usage of causal discovery/inference techniques in the education domain, whereas the previous competition mainly focused on prediction.

The technical challenges of this competition is also novel, e.g. partial observations, selection bias, discovery in groups, etc. These challenges have been largely overlooked by the machine learning community and other competitions, whereas they are ubiquitous in real life. Through this competition we aim to attract some attentions from the community towards these challenges. 

%% file: Website.tex
\paragraph{Website, tutorial and documentation}

The competition is hosted at \url{https://eedi.com/projects/neurips-2022} which serves as a landing page. We will provide the necessary tutorials, documentation, white paper, participants' starter kit at the same time when we release the competition submission site, which will be hosted on Codalab (\url{https://competitions.codalab.org/}). The organizers have rich experience preparing detailed and clear competition materials for the participants; see our white paper from the Diagnostic Questions competition at NeurIPS 2020 \cite{wang2020instructions}.

%% file: Data.tex
\section{Data format}
In this competition, we will provide the participants with two types of data: (1) synthetic data from a simulator; (2) real-world time-series data. For the first part of the competition (i.e.~task 1 and 2), the details of the synthetic data will be explained below. 
\subsection{Synthetic Training Data for tasks 1 and 2}
Directly diving into solving the real-world challenge can be intimidating and difficult to start with. To smooth the participation process, we provide synthetic data from a simple simulator that captures the basic learning mechanisms. 
This type of synthetic dataset avoids several practical challenges, such as missing data, selection bias, etc. An example of the synthetic data set is shown in Table \ref{tab:syntheticdata}

\begin{table}[h]
\centering
\captionsetup{width=0.9\linewidth}
\caption{An example of synthetic data set format.}
\begin{tabular}{@{}lcccc@{}}
\toprule
{\bf student\_id} & {\bf bot\_action} & {\bf construct\_0}           & {\bf construct\_1}           & {\bf...} \\ \midrule
0           & 3                               & 0.37  & 0.56 & ... \\
0           & 2                               & 0.43  & 0.72 & ... \\
...         & ...                             &   ...                 &      ...              & ... \\
25          & 15                              & 0.38 & 0.56  & ... \\
25          & 1                               & 0.66 & 0.94  & ... \\
\bottomrule
\end{tabular}
\label{tab:syntheticdata}
\end{table}
This synthetic dataset will be used as the training data for both task 1 and 2. 
The first column, \texttt{student\_id}, specifies the student index. The rows with the same \texttt{student\_id} forms a time series containing the learning path history. The first row of \texttt{student\_id} represents the starting time of the learning path. \texttt{bot\_action} specifies the construct index the bot assigns to the student to learn at that time. The possible assignment is $[0, \ldots,\text{number of construct}-1]$. \texttt{construct\_*} is a value between $0$ and $1$, representing the probability of correctly answering a multiple choice question related to that construct. Since it is difficult to directly measure the student's knowledge of a particular construct, this metric can be regarded as a surrogate for construct knowledge. 

For task 1 and 2, we will provide \textbf{5} datasets each for local development, public leaderboard and private leaderboard. For local development in task 1, we also provide the ground truth relations between constructs, in the form of adjacency matrix, to help the model development. 

\paragraph{Adjacency matrix format}
For both submissions and local development, the construct relations for each dataset are represented by the binary adjacency matrix in the format of matrix \ref{eq: adjacency matrix}. The size of the matrix is \texttt{num\_construct $\times$ num\_construct}. Specifically, \texttt{adj\_matrix[i,j]$=1$} represents \texttt{construct\_i} is a prerequisite for \texttt{construct\_j}, and $0$ means there is no direct relation between them.
\begin{equation} \begin{bmatrix}
0 & 1 & ... & 0 & 1 \\
0 & 0 & ... & 1 & 1 \\
0 & 0 & ... & 0 & 1 \\
1 & 0 & ... & 0 & 1 \\
0 & 0 & ... & 0 & 0 
\end{bmatrix}  
\label{eq: adjacency matrix}
\end{equation}

\paragraph{Dataset statistics} For each dataset, it contains $100$ students and $400$ simulation time for each student. The total number of construct is $50$. Thus, this will contain $40000$ observations, rendering the manual analysis impossible. 

\paragraph{Supplied file} The data folder for local development, public leaderboard and private leaderboard are called \texttt{Task\_1\_data\_local\_dev\_csv}, \texttt{Task\_1\_data\_public\_csv} and \texttt{Task\_1\_data\_private\_csv}, respectively. Inside each folder, it contains 5 sub-folders with name \texttt{dataset\_0}, $\ldots$, \texttt{dataset\_4}, corresponding to different datasets. Inside, it has \texttt{train.csv} file, containing the training data in the format of Table \ref{tab:syntheticdata}. In particular, for \texttt{Task\_1\_data\_local\_dev\_csv}, it contains an additional file \texttt{adj\_matrix.npy}, representing the ground truth relations between construct. It is an \texttt{ndarray} with shape \texttt{[5, 50, 50]}, where the first dimension represents the dataset number, and the remaining part is a binary $50\times 50$ adjacency matrix in the format of matrix \ref{eq: adjacency matrix}. 

\subsection{Intervention Data for task 2}
In order to estimate the \textit{conditional average treatment effect} (CATE), we provide 10 queries for each dataset in Task 1. The set of queries is represented by a \texttt{list} of \texttt{dictionaries}, each \texttt{dictionary} contains the following:
\begin{itemize}
    \item \texttt{conditioning}: The conditioning sample for estimating CATE. It has the shape \texttt{[conditioning\_length, num\_construct+1]}, where the $+1$ is the \texttt{bot\_action}. This data has the same format as Table \ref{tab:syntheticdata} without the student index. This conditioning sample can be regarded as a learning path of a simulated student until the current time. 
    \item \texttt{intervention}: A list with shape \texttt{[1, num\_construct+1]}, specifying the intervened variables and value. The non-intervened variable will be \texttt{NaN}. 
    Due to the formulation of the synthetic data, the only possible intervention variable is the \texttt{bot\_action} (i.e. the first variable in the list). Thus, an example of this intervention is \texttt{[[25,NaN,$\ldots$, NaN]]}, meaning that we intervene the \texttt{bot\_action} to assign construct \texttt{25} to the \textbf{simulated} student mentioned in \texttt{conditioning} to learn at the current time. 
    \item \texttt{reference}: It has the same format as \texttt{intervention}. The only difference is the value for the \texttt{bot\_action}. 
    \item \texttt{effect\_mask}: A boolean list with shape \texttt{[3, num\_construct+1]}. If \texttt{list[i,j]=True}, it means the target variable for the treatment effect estimation is variable \texttt{j} (note that variable number is not the same as construct number due to the variable \texttt{bot\_action}) at the future ahead time \texttt{i}. For task 2, we fixed the future ahead time to be always \texttt{2} and there can only be one \texttt{True} value in the list. For example, \texttt{list[2,j]=True} means we want to estimate the effect of the intervention/reference specified in \texttt{intervention}/\texttt{reference} to construct \texttt{j-1} at \texttt{2} time ahead for the simulated student in \texttt{conditioning}. 
\end{itemize}
\paragraph{Text file} We also provide a human readable \texttt{.txt} files summarising the query for each dataset. Each \texttt{.txt} file contains the following:
\begin{itemize}
    \item \texttt{CATE number}: This is the index of the query. The total number of queries is \texttt{$10$} for each dataset.
    \item \texttt{Conditioning\_length}: It summarises the simulation time for the conditioning sample in the dictionary mentioned above. For example, \texttt{Conditioning\_length:140} means the conditioning sample has a shape \texttt{[141, num\_construct+1]} (since the starting time is 0).
    \item \texttt{Bot intervention}: It specifies the intervention value (i.e. construct id) for \texttt{bot\_action}.
    \item \texttt{Bot reference}: Same as \texttt{Bot intervention}.
    \item \texttt{Effect construct}: The target construct id for CATE computation.
    \item \texttt{Effect time}: It specifies time step ahead of the current time for the target variable. For example, \texttt{Effect time:2} representing the CATE is computed for variable \texttt{Effect construct} at \texttt{2} step ahead of simulated time \texttt{Conditioning\_length}.  
\end{itemize}
\paragraph{CATE estimation format}
As described before, each query will generate a CATE estimation, the CATE estimations for a dataset will be an array with \textbf{10} elements where each element corresponds to a query ordered based on the aforementioned \texttt{.txt} or \texttt{.json} files. Since there are \textbf{5} datasets for local development and public/private leaderboard, the CATE estimations should be stored as a numpy matrix with shape \texttt{[5, 10]}. The order of the first dimension should follow the dataset number. Namely, \texttt{cate\_estimate[0,3]} represents the CATE estimation for dataset \textbf{0} and query number \textbf{3}.
\paragraph{Supplied file}
For task 2, we only provide CATE query files specified in this section. The corresponding training data can be found in task 1 dataset. Similar to task 1, we provide three folders:
\begin{itemize}
    \item \texttt{Task\_2\_data\_local\_dev}: It contains the CATE query files for local model development. The corresponding training data folder is \texttt{Task\_1\_data\_local\_dev\_csv}.
    \item \texttt{Task\_2\_data\_public}: CATE query files for public leaderboard. The corresponding training data folder is \texttt{Task\_1\_data\_public\_csv}.
    \item \texttt{Task\_2\_data\_private}: CATE query files for private leaderboard. The training data folder is \texttt{Task\_1\_data\_private\_csv}.
\end{itemize}
Inside each folder, we have the following files:
\begin{itemize}
    \item \texttt{intervention\_*.txt}: The text file specifies the CATE query. \texttt{*} is the dataset number. 
    \item \texttt{intervention\_*.json}: The json file containing the conditioning samples, intervention, reference and effect variables described in this section. 
    \item \texttt{cate\_estimate.npy}: This is only available for \texttt{Task\_2\_data\_local\_dev}, which contains the ground truth CATE estimations with shape \texttt{[5,10]}. 
\end{itemize}

%% file: DataRealWorld.tex
\subsection{Real-world Data for tasks 3 and 4}

\subsubsection{Training Data}

Filename: \texttt{checkin\_lessons\_checkouts\_training.csv}

This is the raw training data obtained from a real-world online education platform. This platform serves diagnostic questions and lessons to school students (roughly between 11 and 16 years old).

Each diagnostic question is a multiple-choice question with four possible answer choices, exactly one of which is correct. Each lesson covers a single construct and consists of explainer videos and self-marked questions. 

A quiz comprises five ``checkin'' diagnostic questions. After each checkin question students may attempt a lesson, although this is optional if they answered correctly. Following the lesson they must answer a ``checkout'' diagnostic question which is associated to the same construct as the checkin question and lesson.

When a student answers a checkin or checkout question incorrectly they must retry the same question before moving on. We record both attempts.

We use only data from student responses to mathematics content collected between February 1st, 2022 and August 3rd, 2022.  

Table \ref{tab:primaryanswerdata} is an illustration of the data records.

\begin{table}[h]
\vspace{15.0pt}
\centering
\captionsetup{width=0.9\linewidth}
\vspace{-25pt}
\caption{Primary answer data.}
\vspace{5pt}
\begin{tabular}{@{}ccccccc@{}}
\toprule
{\bf QuizSessionId} 
& {\bf AnswerId}
& {\bf UserId} 
& {\bf QuizId} 
& {\bf QuestionId} 
& {\bf IsCorrect} 
& {\bf AnswerValue} \\
\midrule
8 & 57 & 5 & 232950 & 131432 & 0 & 2\\
8 & 58 & 5 & 232950 & 131432 & 0 & 3\\
8 &	None & 5 & 232950 & 131432 & None & None\\
8 & 59 & 5 & 232950 & 133665 & 1 & 4\\
8 & 60 & 5 & 232950 & 131433 & 1 & 1 \\
\bottomrule
\end{tabular}
\label{tab:primaryanswerdata}
\end{table}

\begin{table}[h]
\vspace{15.0pt}
\centering
\captionsetup{width=0.9\linewidth}
\vspace{-20pt}
\begin{tabular}{@{}ccccc@{}}
\toprule
{\bf CorrectAnswer}
& {\bf QuestionSequence}
& {\bf ConstructId}
& {\bf Type}
& {\bf Timestamp} \\
\midrule
4 & 2 & 433 & Checkin & \dots 06:15:01\\
4 & 2 & 433 & CheckinRetry & \dots 06:16:18\\
None & 2 & 433 & Lesson & \dots 06:26:19\\
4 & 2 & 433 & Checkout & \dots 06:27:03\\
1 & 3 & 427 & Checkin & \dots 06:30:41\\
\bottomrule
\end{tabular}
\label{tab:primaryanswerdatcontinued}
\end{table}

Each row in the table represents an event during a quiz with an associated \texttt{Timestamp}. Either this is for an attempt at a question or a lesson. The \texttt{Type} for an attempt at a question ''Checkin'', ''CheckinRetry'', ''Checkout'', or ''CheckoutRetry''. The \texttt{Type} for attempting a lesson is ''Lesson''.

The \texttt{QuizSessionId} identifies an attempt at a quiz by a user, which are in turn identified by a \texttt{QuizId} and \texttt{UserId} respectively. The questions within each quiz are identified by their \texttt{QuestionId} and each is linked to a single construct which is identified by the \texttt{ConstructId}. The user's answer to a question is identified by the \texttt{AnswerId}.

\texttt{IsCorrect} represents whether a student has answered a question correctly (1) or incorrectly (0). There are four possible answers to each question, one of which is correct. The student's answer is recorded as \texttt{AnswerValue} (1 to 4) and the correct option as \texttt{CorrectAnswer} (1 to 4). 

The \texttt{QuestionSequence} (1 to 5) records the position of the questions within the quiz and allows us to link checkin questions, lessons, and checkout questions. 

The data contains both incomplete and complete quiz attempts by students. In total there are over 65,000 quiz attempts from 6400 students. These involve over 470,000 answers to diagnostic questions and 37,000 attempts at lessons.

\subsubsection{Metadata}

\paragraph{Topic Pathway Metadata} (\texttt{topic\_pathway\_metadata.csv}) Eedi's topic pathway is a sequence of topics in an order which has been recommended by a team of mathematics teachers. Each row in the table represents a pair of questions (\texttt{CheckinQuestionId} and \texttt{CheckoutQuestionId} within a topic quiz (\texttt{QuizId}). The position of questions within a quiz is given by the \texttt{QuestionSequence}, and the position of the quiz within the topic pathway is determined by the \texttt{QuizSequence}. The quizzes are grouped into \texttt{Level}s which cover different \texttt{YearGroup}s. Each question pair covers a specific \texttt{ConstructId} which is linked itself to a single \texttt{SubjectId}. Questions can be linked directly to multiple subjects and we record this as a comma-separated list in \texttt{QuestionSubjectIds}.

\paragraph{Subject Metadata} (\texttt{subject\_metadata.csv}) Each subject is identified by a \texttt{SubjectId} and has an associated \texttt{Name}. The table is self-referencing, which means the \texttt{ParentId} of one subject is a \texttt{SubjectId} of another subject. The \texttt{Level} tells you how many degrees of separation a subject is from the top-level subject, ''Maths''.

\paragraph{Student Metadata} (\texttt{student\_metadata.csv}) Each user is identified by a unique \texttt{UserId}. The \texttt{Gender} is either ''male'', ''female'', ''other'' or ''unspecified''. Instead of a date of birth, we record a \texttt{MonthOfBirth} which is the 1st day of the month in which they were born.  The \texttt{YearGroup} is self-reported by the user and not calculated from the date of birth. These year groups follow the UK numbering system: \url{https://www.gov.uk/national-curriculum}.

In the UK, schools receive addition funding for disadvantaged students, which is known as ''pupil premium''. Whether a user is a pupil premium student (1) or is not (0) is recorded in the \texttt{IsPupilPremium} column. 

When demographic information is unknown it is left blank in this table.

This metadata has been shared solely for the purpose of this competition and must not be used for any other purpose. Data must not be shared with anyone outside of the competition.

\subsubsection{Test Data}
\label{subsubsec: AB experiment data}
For real-world evaluation we have collected additional intervention data. We have run several A/B tests to establish the relationships between completing a lesson on a given construct and performance in questions on other constructs. These tests were conducted by splitting students randomly into treatment and control groups. Both groups were given a a quiz of five check-in questions from some constructs. The treatment group were then given a lesson on a related construct, whilst the control group were given a lesson on an unrelated construct. These relationships were defined by domain experts.  Both groups were then given a quiz of five check-out questions on the same constructs as the check-in questions.

Filename: \texttt{constructs\_input\_test.csv}

The file contains a construct list for task 3, which the participants are required to infer an adjacency matrix indicating the presence and direction of causal relationships.

Filename: \texttt{construct\_experiments\_input\_test.csv}

This is the questionnaire for task 4, where each row contains an query that the participants need to compute CATE for. 
Table \ref{tab:task4inputtest} provides an example of this file. Each row describes an experiment which we have run. Each experiment uniquely defined by the pair \texttt{(Experiment, QuestionConstructId)}. The construct being tested by the check-out questions is \texttt{QuestionConstructId} (i.e.~target construct), and the constructs being taught in the treatment and control lessons are \texttt{TreatmentLessonConstructId} and \texttt{ControlLessonConstructId} respectively. Finally the \texttt{Year} indicate the year group information of the participated students. 

\begin{table}[h]
\vspace{15.0pt}
\centering
\captionsetup{width=0.9\linewidth}
\vspace{-25pt}
\caption{Task 4 input test data.}
\vspace{5pt}
\begin{tabular}{@{}ccccc@{}}
\toprule
{\bf Experiment} 
& {\bf QuestionConstructId}
& {\bf TreatmentConstructId} 
& {\bf ControlConstructId} 
& {\bf Year} \\
\midrule
7\_1 & 471 & 469 & 2930 & 7 \\
10\_1 & 2034 & 2028 & 628 & 10\\
\bottomrule
\end{tabular}
\label{tab:task4inputtest}
\end{table}

%% file: Task.tex
\section{Task Details}
\label{sec: Task}
\input{task_1}
\input{Task_2}

\input{task_3}
\input{Task_4}

%% file: task_1.tex
\subsection{Task 1: Relationship discovery for constructs over time using synthetic time-series data}
\label{subsec:task1}

The relationship among different constructs is key for us to set the learning path for students. Currently, in education, the relationships are determined by teachers and these vary among different teachers depending on their experience. In this task, we would like to discover the construct relationships using \texttt{complete} synthetic time-series data. The causal discovery results will be evaluated against the ground-truth causal graph. Apart from the test datasets, we provide additional synthetic datasets for self-evaluations (\texttt{Task\_1\_data\_local\_dev\_csv}). All datasets will use different construct relationships.

Although the nature of the training dataset is temporal, we are interested in the time-aggregated causal relationships among constructs for each dataset. This aggregated relationship will be represented in the format of binary adjacency matrix as matrix \ref{eq: adjacency matrix}. 

\subsubsection{Evaluation Metric}
Participants are asked to submit a temporally aggregated adjacency matrices for inter-construct relations for each dataset (5 datasets in total). Specifically, the ground truth temporal adjancency matrix for $n$ constructs $A$ with shape \texttt{[lag+1, n+1, n+1]} is aggregated by cropping the \texttt{bot\_action} from the adjacency matrix and performing a "logical or" over all lag adjacency matrices:
\begin{equation}
A_{agg} = \left(\sum_t A_{\{t, 1:n+1, 1:n+1\}}\right) \geq 0.
\end{equation}
\textbf{Note: we do not want to limit possible method choices for this task. It is not necessary to first estimate the temporal adjacency matrix, followed by aggregation. One can use the method that directly infers the aggregated construct adjacency matrix. What we described bove is how we generate the ground truth aggregated adjacency matrix.}

Given the submitted aggregated adjacency matrix $A_{agg}$ and the true aggregated adjacency matrix $\hat{A}_{agg}$, we calculate the $F_1$-score for correctly classifying each (non-diagonal) variable ($i$) - variable ($j$) relationship $r_{ij}$ as one of 4 categories: (0) $i \nleftrightarrow j$; (1) $i \rightarrow j$; (2) $i \leftarrow j$; (3) $i \leftrightarrow j$. Given the true relationships $\hat{r}_{ij}$ and derived submitted relationships $r_{ij}$ we calculate the metric as follows,
\begin{align}
    \mathrm{recall}^{(d)} &= \frac{\sum_{i=1}^{n} \sum_{j=i}^{n} (r_{ij} = \hat{r}_{ij}) * (\hat{r}_{ij} \neq 0) }{\sum_{i=1}^{n} \sum_{j=i}^{n} (\hat{r}_{ij} \neq 0)}, \\
    \mathrm{precision}^{(d)} &= \frac{\sum_{i=1}^{n} \sum_{j=i}^{n} (r_{ij} = \hat{r}_{ij}) * (r_{ij} \neq 0) }{\sum_{i=1}^{n} \sum_{j=i}^{n} (r_{ij} \neq 0)}, \\
    F_1^{(d)} &= \frac{2 * \mathrm{recall^{(d)}} * \mathrm{precision^{(d)}}}{\mathrm{recall^{(d)}} + \mathrm{precision^{(d)}}},
\end{align}
where $d$ is the index of the dataset. The final metric is $\bar{F_1} = \frac{1}{5} \sum_{d=1}^{5} F_1^{(d)}$.

%% file: Task_2.tex
\subsection{Task 2: Teaching Effectiveness Inference using Synthetic Time-series Data}
\label{subsec: task 2 details}
Given the limited learning time for students, we would like to provide the student with the learning material that is most helpful for their overall knowledge acquisition. In addition, since each student may have different learning path history, the provided learning material should also take this into consideration. 
Thus, temporal conditional average treatment effect (CATE) is a good measure on how effective learning a particular construct helps another target construct. We use the probability of correctly answering the associated question as a performance indicator for that construct. 
For each synthetic dataset in task 1, we will provide $10$ queries. For each query, we will give
\begin{itemize}
    \item Conditioning sample: It acts as the learning path history of the student up until the current time.
    \item Intervention: It specifies the intervened construct id for the bot to assign. 
    \item Reference: It specifies the reference construct id for the bot to assign.
    \item Effect: It specifies the target construct id where we want to compute the treatment effect for.
    \item Effect time: Since we are interested in the treatment effect in the future, this represents the how many time steps is ahead of the conditional sample. For example, $0$ means the current time (i.e. $1$ step ahead of conditional sample), $1$ means $1$ step ahead in the future (i.e. $2$ step ahead of conditional sample), etc. For this task, the effect time is always set to $2$.
\end{itemize}

The goal is to compute the CATE for each intervention/reference-effect pair (10 for each dataset) conditioned on conditioning sample (i.e. history learning path). Formally, CATE is defined in Eq.\ref{eq: CATE}.

\begin{equation}
\cate(c_I, c_R, Y_{c_t,t+2}, \bm{x}_{0:t-1}) = \mathbb{E}_{p(Y_{c_t,t+2}|\doop(B_t=c_I), \bm{x}_{0:t-1})}[Y_{c_t,t+2}] - \mathbb{E}_{p(Y_{c_t,t+2}|\doop(B_t=c_R), \bm{x}_{0:t-1})}[Y_{c_t,t+2}],
\label{eq: CATE}
\end{equation}
where $c_I$, $c_R$ are intervention and reference construct indices respectively, $B_t$ is the bot action at current time $t$, $Y_{c_t,t+2}$ is the probability of correctly answering the question associated with the target construct $c_t$ at furture time $t+2$, and $\bm{x}_{0:t-1}$ is the conditioning learning path history (i.e.~the conditioning sample). 
\subsubsection{Evaluation Metric}
Participants submit their estimation $\cate_{k,d}'$ for a set of $10$ predetermined queries for each of the dataset in task 1 ($5$ datasets in total). We compare those predictions to the ground truth $\cate^*_{k,d}$ using rooted mean square error (RMSE):
 
\begin{equation}
    \mathrm{RMSE} = \frac{1}{5}\sum_{d=0}^4\sqrt{\frac{1}{10}\sum_{k=0}^9\left(\cate'_{k,d}-\cate^*_{k,d}\right)^2}, 
\end{equation}
where $k$ is the query id and $d$ is the dataset number. 
Submissions with missing the CATE estimation are invalid and not considered. The ranking in public/private leaderboard is based on \textbf{negative RMSE} in an \textbf{descending order}.

%% file: task_3.tex
\subsection{Task 3: Relationship discovery for constructs over time using synthetic time-series data}
\label{subsec:task3}

Similar to task 1 (Section \ref{subsec:task1}), task 3 focuses on causal discovery from real-world data. The major difference compared to task 1 is the following: The training data in task 4 are no longer \textbf{complete}. Namely, at each timestamp, we only have the access to a binary value ($0$ or $1$) indicating whether a particular question is correctly answered or not, instead of the overall question accuracy (i.e.~a good surrogate for construct knowledge) for synthetic data. 

Additionally, we do not have access to the full true adjacency matrix. Instead we were only able to identify specific edges through experiments. Therefore, we provide a list of constructs for which the adjacency matrix should be submitted. The file \texttt{constructs\_input\_test.csv} contains a list of the requested constructs.

Although the nature of the training dataset is temporal, we are interested in the time-aggregated causal relationships among constructs for each dataset. This aggregated relationship will be represented in the format of binary adjacency matrix as matrix \ref{eq: adjacency matrix}. 

\subsubsection{Evaluation Metric}
Participants are asked to submit a temporally aggregated adjacency matrices for inter-construct relations similar to Task 1. We're using the same metrics as in Task 1. However, not all edges between all constructs are known and we calculate the metrics only on experimentally validated edges. This set of edge pairs will not be released to the participants. 

%% file: Task_4.tex
\subsection{Task 4: Teaching Effectiveness Inference using Real-world Data}
\label{subsec: task 4 details}
With a similar context as task 2 (Section \ref{subsec: task 2 details}), task 4 focuses on the CATE estimation with real-world data. The major differences compared to task 2 are the following:
\begin{enumerate}
    \item The training data in task 4 are no longer \textbf{complete}. Namely, at each timestamp, we only have the access to a binary value ($0$ or $1$) indicating whether a particular question is correctly answered or not, instead of the overall question accuracy (i.e.~a good surrogate for construct knowledge) for synthetic data. 
    \item The CATE definition for task 4 is a little different compared to task 2, since we only access to a questionnaire (i.e.~\texttt{construct\_experiments\_input\_test.csv}) containing the construct pairs we want to compute CATE for, instead of the actual conditioning samples and effect time in task 2. 
\end{enumerate}
In the questionnaire, each entry contains the following:
\begin{itemize}
    \item \texttt{Experiment}: This specifies experiment ID.
    \item \texttt{QuestionConstructID}: This is the target construct ID we need to compute CATE for.
    \item \texttt{TreatmentLessonConstructId}: This is the intervention construct ID.
    \item \texttt{ControlLessonConstructId}: This is the reference construct ID.
    \item \texttt{Year}: This specifies at which year group this experiment is conduced to. 
\end{itemize}
The goal is to compute the following CATE for each of the entry in the questionnaire:
\begin{equation}
    \cate(c_I, c_R, Y_{c_t}, T) = \mathbb{E}_{p(Y_{c_t}|\doop(B=c_I), T)}[Y_{c_t}] - \mathbb{E}_{p(Y_{c_t}|\doop(B=c_R), T)}[Y_{c_t}],
\label{eq: CATE task 4}
\end{equation}
where $T$ is the year group information, $Y_{c_t}$ is the averaged question accuracy for the target construct, $c_I$, $c_R$ and $B$ have the same definitions as task 2.
\subsubsection{Evaluation Metric}
\label{subsubsec: evaluation metric task 4}
Participants need to submit their estimated CATE for all entries in the questionnaire. We compare those predictions to the ground truth CATE obtained from AB experiment (refer to Section \ref{subsubsec: AB experiment data} for details) using RMSE:
\begin{equation}
    \mathrm{RMSE} = \sqrt{\frac{1}{N}\sum_{k=1}^N\left(\cate'_k-\cate^*_k\right)^2},
    \label{eq: RMSE task 4}
\end{equation}
where $N$ is the number of queries in the questionnaire, $\cate'_k$ is the predicted CATE for query $k$ and $\cate^*_k$ is the ground truth.

%% file: Submission_protocol.tex
\section{Submission Protocol}
Task 1, 2, 3 and 4 contain two phases: a public evaluation phase and private evaluation phase. Results in the public evaluation phase are displayed on a public leaderboard allowing participants to see their performances compared to others. Results for private evaluation phase are hidden until the end of the competition. \textbf{Important: For each task, participants must submit to both the public and private evaluation phases separately. Submissions made solely to the public evaluation phase will not be used in the final judgement of the competition. It is the participants’ responsibility to make sure their submission to the private phase of each task represent their best results. The submission file should be a single \texttt{zip} file, containing a single \texttt{npy} file inside. The name of the \texttt{npy} file should have a specific name defined below. On the other hand, for local evaluation (task 1 and 2 only), only \texttt{npy} files are needed (refer to the evaluation README in the starting kit). }
\subsection{Submission for Task 1}
According to the task descriptions in section \ref{subsec:task1}, the adjacency matrices are stored as a numpy array with shape \texttt{[5, 50, 50]}, with the file name \texttt{adj\_matrix.npy}. Alternatively, it is possible to submit probabilistic estimates of the adjacency matrix in the form of samples from the adjacency matrix distribution. This should be submitted in the shape of \texttt{[5, s, 50, 50]}, where $s$ is the number of samples from the distribution over adjacency matrices. Accordingly, the F1 score will be averaged over the batch $s$. 

A template \texttt{npy} file can also be obtained by running the task 1 baseline model. 
\subsection{Submission for Task 2}
According to the task descriptions in section \ref{subsec: task 2 details}, the CATE estimations are stored as a numpy array with shape \texttt{[5, 10]}, with the file name \texttt{cate\_estimate.npy}. This array should not contain any \texttt{NaN} values and its shape has to exactly match the \texttt{[5,10]}. The elements inside the array should be ordered according to the dataset index and query id. For instance, \texttt{cate\_estimate[3,7]=0.45} means the CATE estimation of query $7$ for dataset $3$ is $0.45$. 

A template \texttt{npy} file can also be obtained by running the task 2 baseline model. 

\subsection{Submission for Task 3}
For task 3, the candidates need to submit an adjacency matrix for a subset of constructs listed in \texttt{constructs\_input\_test.csv}, with the file name \texttt{adj\_matrix.npy}.
It should have the shape \texttt{[num\_constructs, num\_constructs]}
Alternatively, similar to task 2, it is possible to submit a probabilistic estimates in the form of samples. It should have shape \texttt{[s,num\_constructs, num\_constructs]}, where \texttt{s} is the number of adjacency samples. The F1 score will be computed for each sample, followed by averaging over them. 

A template \texttt{npy} file can also be obtained by running the task 3 baseline model. 
\paragraph{Important} Since the submitted \texttt{npy} adjacency matrix does not have the labels for the constructs, the order of the constructs is important. The submitted matrix should respect the order in \texttt{constructs\_input\_test.csv}. For example, if \texttt{constructs\_input\_test.csv} contains a list \texttt{[1030,2066,531,3358, ...]}, the \texttt{adj\_matrix[0,3]} should indicate the connection from \texttt{1030} to \texttt{3358}. For public leaderboard, we will not reveal the real score of the submission. However, the rankings of the participants are based on the real score. 

\subsection{Submission for Task 4}
According to the task descriptions in section \ref{subsec: task 4 details}, the predicted CATE should stored as a numpy array with shape \texttt{[num\_query]} and file name \texttt{cate\_estimate.npy}. It should not contain any \texttt{NaN} values and has to exactly math the shape. The order in the numpy array should follow the questionnaire. A template \texttt{npy} file can be obtained by running the task 4 starting kit. 
\paragraph{Important} For the task 4 public leaderboard, we will not reveal the real score of the submission. However, the rankings of the participants are based on the real score.

%% file: starting_kit.tex
\section{Starting Kit}
For easier participation process, we also provide baseline method and evaluation scripts for each task. They can be downloaded at the shared onedrive\footnote{https://1drv.ms/u/s!Ai2oY6KqGQ4Lb7SWI0hP5-geV3c?e=msNdQp}. Our baseline models for all tasks are based on an open sourced repo called \textit{Causica}\footnote{https://github.com/microsoft/causica}, which contains the implementations of several well-known causal discovery and inference algorithms. In order to run the baseline, candidates need to clone the \textit{Causica} repo to their local machine. Detailed instructions on how to run the baseline method will be described in the \texttt{README} file of the starting kit. 

In the following, we will give a general overview of the provided files.
\subsection{Task 1}
The baseline for task 1 uses VARLiNGaM \citep{hyvarinen2010estimation}, which extends the common linear non-Gaussian model to a vector auto-regressive model and performs causal discovery from time-series data. Originally, this model only works with a single time-series. As such, we select the longest time-series from the training data and perform causal discovery on this series.

The code in the \textit{Causica} repository provides a wrapper for a VARLiNGaM implementation and our starting kit provides the code to run the model and prepare the learned causal graphs for submission. They are stored in starting kit folder: \texttt{starting\_kit/task\_1/}. Inside, it contains
\begin{itemize}
    \item \texttt{README.md}: It contains the detailed instructions on how to perform training and causal discovery estimation with VARLiNGaM.
    \item \texttt{dataset\_config.json}: A config specifying that the dataset is a temporal causal dataset.
    \item \texttt{prepare\_submission.py}: This script will gather the output adjacency matrices from the model save files for each dataset and generate the submission-ready \texttt{adj\_matrix.npy} file. 
\end{itemize}
\subsection{Task 2}
The baseline for task 2 is an adaptation of DECI \citep{geffner2022deep}, which was originally developed for static data. To handle the time series, we adopt a method called \textit{fold-time} trick, where we assume the temporal causal graph is stationary \citep{runge2018causal} and place a fixed cropping window on the temporal graph. Any connections outside the cropping window will be ignored. The size of the window is determined by the model \texttt{lag} and equals to $\text{lag}+1$. Thus, the cropped temporal graph can be regarded as a larger static graph with the restriction that the connection arrow cannot go against time. This static graph with restrictions can be easily handled by DECI. Thus, we name this temporal causal model as \textit{Fold-time DECI} (FT-DECI). 

The implementation of FT-DECI is already in \textit{Causica} repo. The starting kit mainly provides the scripts and functions for CATE estimation with FT-DECI. They are stored in starting kit folder: \texttt{starting\_kit/task\_2/}. Inside, it contains
\begin{itemize}
    \item \texttt{configs}: A folder containing necessary config files to train FT-DECI. 
    Specifically, \texttt{model\_ config\_fold\_time\_deci\_competition.json} contains the training and model hyperparameters.
    \item \texttt{README.md}: It contains the detailed instructions on how to perform training and CATE estimation with FT-DECI.
    \item \texttt{task\_2\_cate.py}: The python script to perform CATE estimation for a particular dataset with the trained FT-DECI.
    \item \texttt{task\_2\_util.py}: It contains utility functions for CATE estimations.
    \item \texttt{prepare\_submission.py}: This script will gather the output CATE estimations from \texttt{task\_2\_cate.py} for each dataset and generate submission \texttt{cate\_estimate.npy} file. 
    
\end{itemize}

\subsection{Task 3 and 4}
Since the FT-DECI used in task 2 can perform both causal discovery and inference at the same time, we use FT-DECI as our baseline for both task 3 and 4. In the starting kit, the scripts are stored at \texttt{task\_3\_and\_4/}. Inside, we provide:
\begin{itemize}
    \item \texttt{README.md}: It contains the detailed instructions on how to train FT-DECI and prepare for submission for task 3 and 4.
    \item \texttt{task\_3\_prepare\_submission.py}: This script will gather the output adjacency matrix from the trained FT-DECI and generate the submission-ready \texttt{adj\_matrix.npy} file for task 3. 
    \item \texttt{task\_4\_cate.py}: This is the main script for computing CATE for task 4. It will output a \texttt{cate\_estimate.npy}. After zipping it, this can be used for submission. 
    \item \texttt{task\_4\_process\_data.py}: This is the script to process the raw training data. Note that we made several assumptions and this is just one example on how to handle the raw training data. The candidates should not rely on this script, and it is mainly used for the starting kit. 
    \item \texttt{configs}: A folder containing necessary config files to train FT-DECI.
    \item \texttt{task\_4\_util.py}: This file contains the utility functions for task 4 baseline model. 
\end{itemize}
    
\paragraph{Important} Please ensure that the \texttt{Causica} repo is updated to the newest release before running the baseline. 

%% file: Organisation.tex
\section{Organizational aspects}
\subsection{Protocol}

All competition tasks require the participants to make predictions based on the provided data. This involves predicting the construct relations and conditional average treatment effects. The participants will need to save the predictions into a \texttt{.zip} file to be submitted to the competition website. We intend to use CodaLab, (\url{https://competitions.codalab.org/}) where the submitted result files will be compared to the ground truth. The participant private rankings will be hidden until the competition closes.

The tasks will be released in two stages, with tasks 1 and 2 released in the first stage and tasks 3 and 4 in the second stage. All tasks will remain open until the competition closes, and public submissions can be made up to 15 per day. Unsuccessful submissions due to errors will not count against this total.

In all cases, code submission will be required in order to check that cheating is not taking place. Submissions exhibiting abnormally good performance will be manually checked, and any winning submissions to the competition will be checked by hand once the competition closes. 

All aspects of the engineering required for running the competition (CodaLab infrastructure, compute resources, networking, participant submission of files and code) will be beta tested prior to opening the competition. This will involve testing the setup on different platforms, in different software environments (but restricted to Python environments as the API will be Python based) and in different geographic locations. We will aim to include users from a diverse range of groups, so as to truly represent the diversity amongst the competitors. Particular attention will be paid to low bandwidth issues when interacting with the platform. 

\subsection{Rules}

The rules of the competition are as follows:

\begin{itemize}
\item The submitted results must be generated by a machine learning model. Submissions making hard-coded predictions are forbidden.
\item Users may make use of open source libraries given proper attribution. At the end of the competition, we encourage all code to be open-sourced so the results can be reproduced. 
\item The use of supplementary existing open-source datasets for e.g. pre-training is permitted provided they are credited. The use of private, proprietary datasets is not permitted.
\item The questions images have been shared solely for the purpose of this competition and must not be used for any other purpose. The question images must not printed or shared with anyone outside of the competition.
\item Any parties or individuals affiliated with Eedi or Microsoft Research are not eligible for the prizes, but may participate in the competition.
\item For participants to be eligible for prizes their code must be publicly available.
\end{itemize}

Any instances of cheating, defined as violation of the rules above or any other attempt to circumvent the spirit and intent of the competition, as determined by the organizers, will result in the invalidation of all of the offending team's submissions and a disqualification from the competition.

In order to prevent cheating, all the evaluation data will be kept completely inaccessible to the participants during the competition. The aforementioned rules on manual review of the submissions also aim to prevent cheating.

\subsection{Schedule and readiness}

\textbf{June 27, 2022: } Tasks 1 and 2 released.

\textbf{August 05, 2022: } Tasks 3 and 4 released.

\textbf{October 15, 2022: } Final submission deadline for all tasks.

\textbf{November 15, 2022: } Results announced, private leaderboards revealed, prize-winners notified.

Submissions will be automatically evaluated on all tasks at time of submission. There is no live/demonstration component to the competition.





%% file: Resources.tex
\section{Resources}

\noindent
The organizing team consists of experts with a range of different backgrounds from both industry and academia.

\begin{itemize}

\item\textbf{Craig Barton} has been involved in teaching maths for over 15 years. He is the Head of Education at Eedi, the TES Maths Adviser, the author of the best-selling books “How I wish I’d taught maths” and "Reflect, Expect, Check, Explain", and Visiting Fellow at the Mathematics Education Centre at the University of Loughborough. (Education domain expert)

\item\textbf{Wenbo Gong} is a Researcher at the Machine intelligence group at Microsoft Research Cambridge (MSRC), UK. Currently, he is a member of project Azua on efficient decision making in MSRC. He is interested in combining the deep learning approaches with causal machine learning to improve efficient decision making and model robustness for real-world impact. (Main coordinator, synthetic data provider, baseline method provider, evaluator)

\item\textbf{Nick Pawlowski} is a Senior Researcher at the Machine intelligence  group at Microsoft Research Cambridge (MSRC), UK. He is a core member of project Azua on efficient decision making in MSRC. He is interested in causal machine learning for robust and reliable decision making for application with business and social impact. (Synthetic data provider, baseline method provider, evaluator)

\item\textbf{Digory Smith} is a Data Scientist at Eedi. He has an MSci in Natural Science from the University of Cambridge and has worked in the education technology sector for the past 7 years. (Coordinator, platform administrator, beta tester, real-world data provider)

\item\textbf{Joel Jennings} is a Senior Research software development engineer at the Machine intelligence group at Microsoft Research Cambridge (MSRC), UK. He is a core member of project Azua on efficient decision making. He has research and engineering experiences in various area of machine learning, including deep learning, causality, Reinforcement learning, etc.

\item\textbf{Zichao Wang} is a Ph.D. student in Electrical and Computer Engineering at Rice University. His research focuses on developing machine learning methods for optimizing how humans learn, acquire, and apply new knowledge. He was the lead organizer of the Diagnostic Questions educational data mining competition at NeurIPS 2020. (Coordinator, platform administrator, beta tester, evaluator)

\item\textbf{Simon Woodhead} is Head of Research at Eedi and holds a PhD in Statistics from the University of Bristol. He has worked in the educational technology sector for 16 years, creating award-winning edtech solutions through innovative data capture, analysis and visualisation. (Coordinator, beta tester, real-world data provider)

\item\textbf{Cheng Zhang} is a Principal Researcher at the Machine intelligence group at Microsoft Research Cambridge (MSRC), UK. Currently, she leads the project Azua on efficient decision making in MSRC. She is interested in both machine learning theory, including Bayesian deep learning, approximate inference and causality for efficient decision making, as well as various machine learning applications with business and social impact. (Coordinator, Advisor, Machine learning domain expert)

\end{itemize}